Short Paper

# Smart Water Irrigation for Rice Farming through the Internet of Things

Regie Binayao
College of Computer Studies, Northern Bukidnon State College, Philippines
regiebinayao@gmail.com

Paul Vincent Mantua
College of Computer Studies, Northern Bukidnon State College, Philippines
paulvincentmantua@gmail.com

Holy Rose May Namocatcat
College of Computer Studies, Northern Bukidnon State College, Philippines
namocatcathyrosemay@gmail.com

Jade Kachel Klient Seroy
College of Computer Studies, Northern Bukidnon State College, Philippines
jadekachelklientseroy@gmail.com

Phoebe Ruth Alithea Sudaria
College of Computer Studies, Northern Bukidnon State College, Philippines
prabacotot@nbsc.edu.ph
(corresponding author)

Kenn Migan Vincent Gumonan
College of Computer Studies, Cebu Institute of Technology University, Philippines
kennmiganvincent.gumonan@cit.edu
https://orcid.org/0000-0002-4790-6729

Shiela Mae Orozco
College of Computer Studies, Northern Bukidnon State College, Philippines
smmorozco@nbsc.edu.ph

**Abstract**

*Purpose* – This study intends to build smart water irrigation for rice farming using IoT and micro-controller devices with solar panel support. The system demonstrates the capabilities of automated irrigation by reducing physical labor through smart monitoring of the temperature, soil moisture, and humidity using multiple sensors.

*Method* – This paper uses an agile methodology as it is suitable for reiterative operation for the development of the prototype.

*Results* - The mean result for the interpretation of data gathered for the system's adaptability and flexibility is 4.32. This figure shows that the selected farmers who took the survey understood and agreed with the prototype and its capability as an effective functionality to solve the problems in rice farming.

*Conclusion* – The researchers were able to develop a smart water irrigation for rice farming using IoT and micro-controller devices with solar panel support and the respondents also agreed that the Smart water irrigation for rice farming using IoT and micro-controller devices with solar panel support is practical and valuable.

*Recommendations* – Integration of a decision support system: A decision support system that can Analyze data collected from IoT sensors and provide further recommendations. Based on the results, it is also suggested that future researchers use drip irrigation, instead of flood irrigation.

*Research Implications* – Smart water irrigation has the potential to revolutionize agriculture, enhance environmental sustainability, and address pressing global challenges related to water resources and food security. These implications highlight the importance of continued research and innovation in this field.

_______________________________________________________________

# INTRODUCTION

Rice is life, especially for Filipinos and other Asians. Even with the availability of other staple foods such as noodles and bread, rice remains the primary choice for staple food in the Philippines (National Nutrition Council, 2020). Rice farming in the Philippines is plagued with problems both from natural and agricultural adversities. Common problems encountered by rice farmers were: the high cost of inputs, low price of rice, lack of capital, labor problems, lack of post-harvest facilities, pests and diseases, and irrigation system (Rani et al., 2019). The agriculture industry is crucial to meeting the nation's needs for food resources (Dhruva et al., 2021). According to Keller of the International Water Management Institute, agriculture continuously contends with home, industrial, and environmental users for a finite amount of water. The availability of water to support healthy crops is either insufficient or excessive. Farmers may find it difficult during the dry season, to get, and farmers have to decide how to share the water for the rice fields equally.

The rice fields that are close to the water source benefit considerably since they receive a lot of water, but those that are far from the source struggle to get water supplies. In daily operations associated with farming, watering is the most vital cultural practice, especially in rice cultivation. Smart irrigation system uses the right amount of water in the right place. It lowers effort and saves time. Even weather patterns are detectable by it. Because of this, a battery, rapid charger, solar panels, and sensors have been included in an automatic irrigation system. The dynamic water above and below the surface is detected by soil moisture, temperature, and humidity sensors (Shufian et al., 2021). Farmers will find it easier to worry about their rice fields and encounter fewer issues with rice field irrigation thanks to the Internet of Things (IoT) information on water requirements in rice fields being monitored at a distance (Rawal, 2017).

The goal of the development of Smart Water Irrigation for Rice Farming through the Internet of Things (SWIMPS) is to make automatic water irrigation and smart monitoring of temperature, humidity, and soil moisture to lessen the farmer's manual operation in rice farming and to save input costs that would directly affect the rice agricultural scheme by using renewable energy specifically the integration of solar panel in the system. This study will greatly benefit rice farmers, especially in the acknowledgment of technology in the agriculture of rice farming.

## LITERATURE REVIEW

### *Water Management*

The significance of efficient management of water is essential in agriculture to attain optimum yield. Smart agricultural water technology highlights how managing smart irrigation systems can help alleviate the stress of farmers by allowing remote monitoring and control of rice farming (Abioye et al., 2022). The best possible water resource usage in the precision farming environment can be achieved with the aid of IoT-based smart irrigation management systems. Using sensing of ground factors like soil moisture, soil temperature, and ambient conditions as well as online weather forecast data, a smart system may estimate a field's need for irrigation. The sensing nodes that are involved in environmental and ground sensing take into account soil moisture, air temperature, and crop field relative humidity (Alhasnawi et al., 2020). Moisture sensors embedded in the field continuously monitor water levels, and upon detecting any fluctuations, they send interrupt signals to the microcontroller for prompt response (Yasin et al., 2019). The irrigation methods that were reviewed support crop growth with the help of sensors like temperature, humidity, soil moisture, and air moisture to provide useful information to the user's decision-making. Internet of Things-based smart irrigation leads to higher crop development, effective water management, and remote access as compared to conventional irrigation systems (Karunakanth et al., 2018; Abdikadir et al., 2023; Deshmukh et al., 2020). The TEROS 12 soil moisture sensor is integrated with a Digi XBee wireless module to gather data on volumetric water content, temperature, and electrical conductivity. These measurements are securely transmitted via an IP gateway to the cloud for analysis and decision-making. This intelligent water irrigation system incorporates a drip irrigation unit, a wireless communication component, and a sensor network to efficiently manage water distribution based on cloud-based control (Guevara et al., 2020).

## Smart Farming

A user-friendly must also be provided that allows users to monitor the field through an application. Additionally, it autonomously regulates and monitors soil irrigation (Ezenwobodo & Samuel, 2022). Utilizing the website for communication and monitoring authorizes users to interact with the sensors instantly. This proves beneficial for users engaging with the microcontrollers from any location, reducing power consumption and enhancing the system's lifespan, all achieved with a relatively modest investment (Sanjana Pandey, 2021). A strategy that aims to enable remote accessibility to the system, ensuring that farmers have constant information and control over their fields 24/7 throughout the entire year was configured through a Renesas microcontroller, which manages the transmission and reception of data through a GSM module (Kumari & Singh, 2021). Utilization of the BLYNK application on a mobile platform offers a streamlined approach to advancing Internet of Things (IoT) projects. This versatile platform not only simplifies the development process but also empowers users with the ability to create customized interfaces tailored to their specific IoT applications. In the context of an irrigation system, BLYNK plays a pivotal role by providing a user-friendly interface on mobile devices. The BLYNK application enables users to effortlessly monitor and control their irrigation systems. Furthermore, BLYNK supports compatibility with a diverse range of IoT hardware platforms and communication protocols, offering flexibility in device integration. This adaptability makes it an ideal choice for those seeking to implement IoT solutions across different devices and environments (Rathore et al., 2023).

## Internet of Things

Through automation and a reduction in human effort, IoT can improve the efficiency of agricultural and farming activities (Kondur et al., 2021; Velmurugan et al., 2020). The exact use of this technology in conjunction with the Internet of Things (IoT) is intended to be the technology that helps farmers improve their standard of life through high productivity and profit. The need for human involvement can be significantly reduced through the implementation of this technology. With all the positive things involved the challenges of components' high costs, lack of internet access, and a lack of application knowledge among farmers, there are still restrictions on the conversion and adoption of smart farms today (Virk et al., 2020; Rafique et al., 2021). Another aspect of smart water irrigation that needs to be addressed is power sources. Because of this, a battery, rapid charger, and solar panels must be included in an automatic irrigation system (Shufian et al., 2021). All of the components concerned from the Arduino and to all of its sensors and other components, need a stable power source to make things work together to achieve the intended result. Emphasis on power generation and consumption must always be put into consideration especially those that come from renewable energy sources (Sudharshan et al., 2019).

The researchers reviewed the existing studies to derive all the functionalities needed for the SWIMP system to address the gaps of the previous research. All the areas for robust irrigation management system through the Internet of Things was not addressed by the relevant literature, but with the Smart Water Irrigation System, all the functionalities are uniquely possessed. Based on the reviewed related literature, there is a need for the development of Irrigation Systems that utilize renewable energy sources such as the integration of Solar panels. When there is a surplus of solar energy, the controller enables the system to charge the battery. In daily operations associated with farming, watering is the most vital cultural practice, especially in rice cultivation. The use of humidity, temperature, and moisture sensors is also a must to help optimize irrigation operations by providing real-time data of soil condition notification to the mobile application in the farmer's smartphones. The proponents also utilized Node-mcu instead of the Arduino microcontroller that was used by the relevant studies, because it offers a convenient cost-effective solution for developers who need wireless fidelity connectivity.

## METHODOLOGY

### *Research Design Approach*

The researchers utilized agile methodology as shown in Figure 1, which employs iteration methods, a series of stages that are repeated with each cycle being utilized to modify and enhance the system (Eby, 2017). It provides a thorough overview of how the hardware functions and enables the project's developers to adjust between developing the hardware and the software and, make necessary changes for improvements.

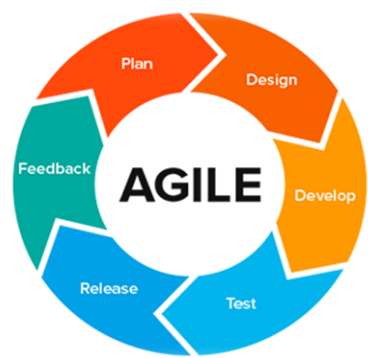

Figure 1. Agile Methodology

### *Planning Phase*

The researchers created a Gantt chart that outlines the order of the activities conducted during the study's development. It began with information collection, which was a crucial step in conducting the study to determine the aim and advantages of the project. To determine potential outcomes and establish objectives for the system output, the researchers deliberated the data gathered to attain the necessary functionalities to be integrated into the system.

## *Functional Requirements*

The following Functional Requirements are the major components used in Smart Water Irrigation IoT.

1. The Solar Panel is added to sustain an energy supply of 12 volts to the entire system.
2. TP4056 Lithium-Ion battery is a backup battery supply that protects the system from power outages and power fluctuation.
3. ESP8266 module is a Wireless fidelity module integrated into the microcontrollers for the system to connect to the Internet.
4. DC water pumps are included to circulate, pressurize, and emulsify water.
5. DHT11 is a basic, ultra-low-cost digital temperature and humidity sensor. It measures the humidity in the air using a thermistor and a capacitive humidity sensor, and it outputs a digital signal on the data pin.
6. Soil moisture sensors measure the water content in the soil and can be used to estimate the amount of stored water in the soil horizon.
7. BLYNK is an IoT platform for iOS or Android smartphones that is used to control Arduino, Raspberry Pi, and NodeMCU through the Internet. This application is used to create a graphical interface or human-machine interface (HMI) by compiling and providing the appropriate address on the available widgets.

## *Design Phase*

Designing the system is an important part of the process, where the researchers start to specify what to create in the making phase based on the information the researchers gathered in the planning phase. In this phase, the researchers prepared the specifications and design of the system hardware, some of the system's hardware components specifications were based on the following research: "AQUAMAG: Smart Water Quality Monitoring through Internet of Things" (Tubio et., Al, 2023) and "GULP: Solar-Powered Smart Garbage Segregation Bins with SMS Notification and Machine Learning Image Processing" (Sigongan et al., 2023).

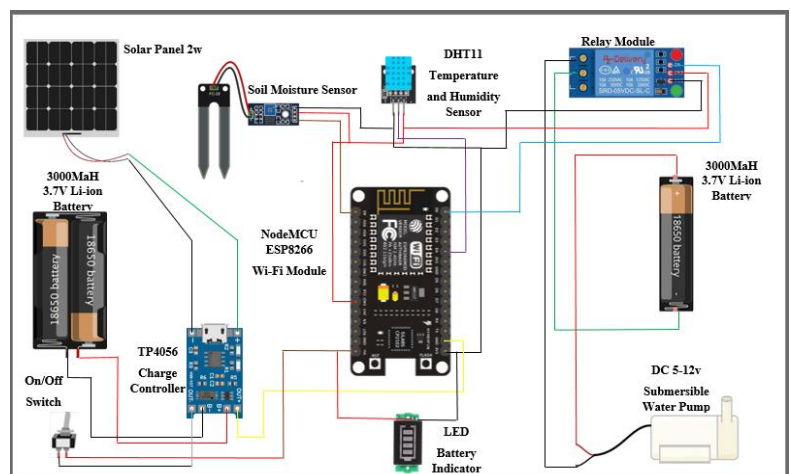

Figure 2. Components Connections of the Device

*Development Phase*

The researchers developed the mobile User Interface using an open-source Blynk application and programmed the microcontroller using the Arduino IDE Application. The process of the system initially starts when the soil moisture sensor senses low detection of water content in the soil. The system will then enable the water pump so that it will release an ample amount of water enough to maintain the soil moisture level. The current temperature and humidity measured by the DHT11 sensor will be automatically shown in the I2C Oled display.

The most popular IDE for combining hardware and software components in a system is the Arduino IDE. It is used to develop the Smart Water Irrigation IoT system using C++ language, code the system's ESP8266 microprocessor to meet the requirements of the system specification, and display all of the application's features with the aid of the open-source BLYNK IOT application, which focuses on information of things. This microprocessor will serve as the brain of the system, in which it will constantly receive data from the sensors and trigger an output module if certain parameters are met.

*Testing and Release*

The researchers must execute a test of their project to check the system's operation, ensure that each component works as intended, and identify any weaknesses or gaps in the system. After the researchers tested the product and the product passed each testing phase, the product was ready to use. This means that the product is ready to be applied in a real environment by the end users of the system. To completely access the system, the end users must install the Smart Water Irrigation IoT mobile application on their respective phones, after installing the application the user must place the soil moisture component in a rice plant or soil, and place the pipe that connects the Dc water pump to the water reservoir to maintain the water content of the soil.

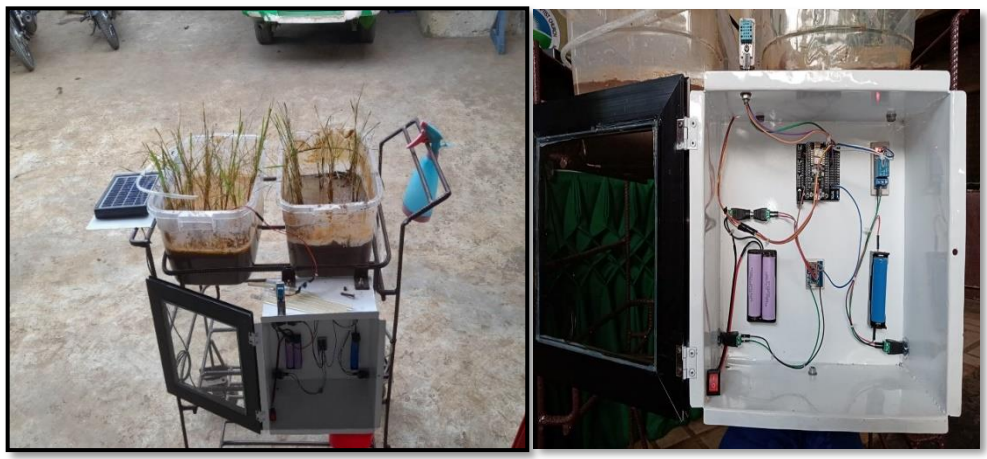

Figure 3. Hardware System in Steel Container Frame with the sample Rice Plant

Figure 3 shows the enclosed container for the System Modules made from Glass and Aluminum material. The system includes the following components ESP8266, Relay Module, Li-ion 3.7v Batteries, TP4056 Charge Controller, Wire Adapters, Solar Panel, DHT11 Sensor, Soil Moisture Sensor, and Plastic Container that contains the Water Pump.

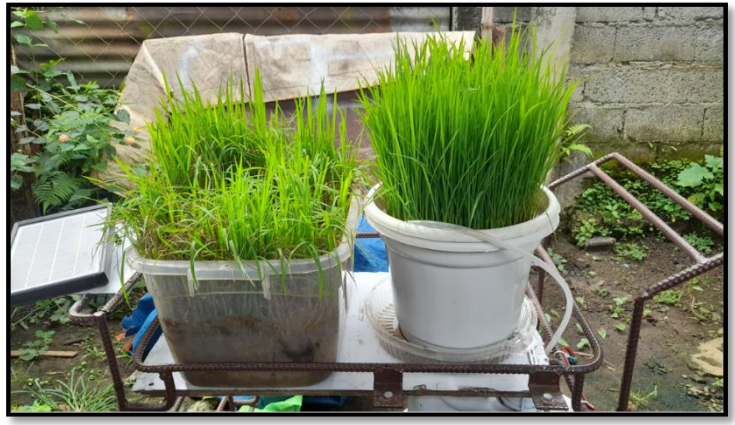

Figure 4. 2 Nursery Rice Seedlings in Pot Containers with the System

Figure 4 shows the actual 2 nursery rice seedlings in pot containers with the Smart Water Irrigation IoT System. After a month of implementing the system to the rice plants' growth procedure. The outcome is that the rice plants are healthy because they yield enough nutrients to grow and the water content of the soil is well balanced.

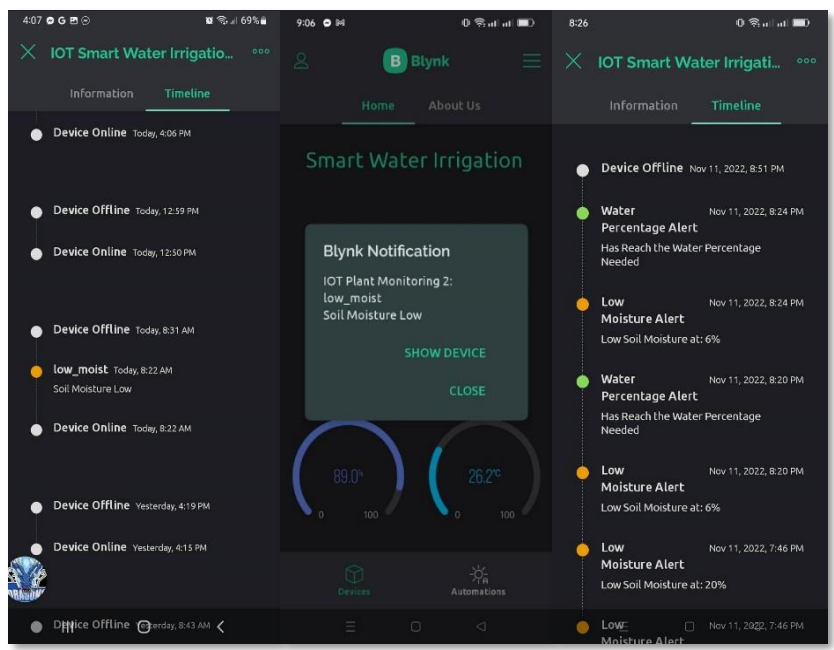

Figure 5. Timeline and Low Moist Notification Displays

Figure 5 shows the Timeline displays of the system. The Application displays notification whenever the Soil Moisture Sensor detects the Low content of water in the soil, it also displays the current temperature and humidity of the rice crop tested with the system.

## Feedback Phase

Feedback is the final phase of the agile development cycle. After the researchers have evaluated and tested the System and have met the required outcomes, the system is now ready to be presented. During the testing and evaluation of the end users, the rice farmers, the survey questionnaires were distributed to them, and users and clients tested the system as part of the evaluation process.

## RESULTS

The system evaluation portion determines the feedback of the personnel and ensures the quality of the final output. ISO/IEC 25010:2011 standards are used for the system evaluation given to the farmer participants. The researchers let the twenty-eight (28) respondents try the system and gathered their feedback by letting them answer the research evaluation questionnaire prepared. The researchers used the Likert scale below in interpreting the data gathered.

The Likert Scale is used to measure people's opinions on a subject matter or specific topic. A 5-point scale was used. The average weighted mean was computed and interpreted as (5) 4.20-5.00: Excellent; (4) 3.40-4.19: Very Good; (3) 2.60-3.39: Good; (2) 1.80-2.59: Fair; and (1) 1.00-1.79: Poor. The results are presented in the succeeding sections of this paper. The researchers used the Likert Scale to interpret the result and overall average of the ISO/IEC 25010:2011 evaluation.

Table 1. ISO/IEC 25010:2011 Evaluation Over Result

| Characteristics | Mean | Verbal Interpretation |
|---|---|---|
| Functional Sustainability | 4.50 | Excellent |
| Performance Efficiency | 4.22 | Excellent |
| Compatibility | 4.33 | Excellent |
| Usability | 4.13 | Very Good |
| Reliability | 4.13 | Very Good |
| Security | 4.55 | Excellent |
| Maintainability | 4.25 | Excellent |
| Portability | 4.42 | Excellent |
| Overall Weighted Mean | 4.32 | Excellent |

Table 1 shows the Overall results that were supplied for the ISO/IEC 25010.2011 assessment. The findings support the notion that the feedback provided during the presentation itself was constructive. Functional sustainability is rated at 4.50, indicating an excellent level of performance in terms of maintaining functionality over time. Performance efficiency is scored at 4.22, reflecting excellent efficiency in carrying out tasks. Compatibility receives a rating of 4.33, signifying excellent adaptability and integration capabilities with other systems. Usability is assessed at 4.13, marking it as very good and suggesting a high degree of user-friendliness. Similarly, reliability is also rated at 4.13, indicating a very good level of dependability in the system's performance. Security is highly rated at 4.55, indicating an excellent level of protection against unauthorized access and potential threats. Maintainability is scored at 4.25, reflecting the system's excellent ease of maintenance and updates. Portability is rated at 4.42, indicating excellent adaptability for use in various environments. The respondents regarded the Smart Water Irrigation as functional and advantageous to them, giving it an overall average satisfaction rating of 4.32 which was verbally interpreted as "Excellent".

## DISCUSSIONS

The study has achieved a comprehensive compilation of both hardware and software specifications essential for measuring soil moisture, temperature, and humidity. The primary objective was to analyze the collected data, ultimately determining the requisite software and hardware specifications for the system's development. Integration with the open-source application, BLYNK IOT, facilitated the display of device outputs, including soil moisture, temperature, and humidity conditions.

The successful design and development of the system involved the utilization of BLYNK IOT, incorporating monitoring gauges and microcontroller-based components for efficient data collection. To validate the system's capabilities, the researchers conducted thorough testing and evaluation. The focus of this evaluation was on assessing the performance of both the software and hardware components.

In practical application, the system was deployed and tested in the agricultural setting of Rice Farmers in Dicklum, Manolo Fortich, Bukidnon. The obtained results, reflected in the 4.32 overall average mean, interpreted as "Excellent," highlighted the system's commendable performance. This positive rating implies that the system is well-suited for the intended purpose and showcases reliability, usability, and efficiency in providing valuable insights into soil conditions for rice cultivation. The success of the study thus validates the practical utility of the developed system in real-world agricultural scenarios.

## CONCLUSIONS AND RECOMMENDATIONS

Based on the result of Table 2 of the Overall results that were supplied from the ISO/IEC 25010.2011 assessment, it is interpreted that the functionality, suitability, performance efficiency, compatibility, usability, reliability, security, maintainability, and portability rated by the respondents was regarded as "Excellent" having an overall weighted mean rate of 4.32.

In conclusion Smart Water Irrigation IoT System meets the objective of the study in which to develop and design the system, the system also integrated the moisture, humidity, temperature, and soil detection functionalities, designed push notification, developed the BLYNK IoT monitoring app., also developed the system with renewable energy through solar panel, and lastly monitored the growth of rice and help the rice farmers in using the system through user testing through the ISO 25010.2011 the system was commended by the end users and rated excellent.

One of the possible additional features is the installation of a water level indication using a Buzzer so that the user can determine whether the water level is full or empty is one of the potential extra features that the project's supporters could include for the benefit of future researchers. Future researchers are advised to add an extra component, such as an ultrasonic sensor to monitor water levels.

## IMPLICATIONS

The research can help refine and improve smart irrigation systems to optimize water usage in agriculture. This not only conserves water resources but also addresses concerns related to water scarcity and drought conditions. Investigating the impact of different smart irrigation techniques on crop yield and quality is essential. Researchers can explore how precise irrigation can enhance productivity and the nutritional content of crops. Climate change is affecting global weather patterns. The research can help develop smart irrigation systems that can adapt to changing weather conditions and ensure sustainable agricultural practices in the face of climate variability. Exploration of how data analytics and machine learning can be used to analyze the vast amount of data generated by smart irrigation systems.

The research in this area can lead to predictive models for optimized irrigation scheduling. Which is a good avenue for the integration of Precision Agriculture. This investigates how smart irrigation can be integrated with other precision agriculture technologies such as GPS-guided tractors and drones. This interdisciplinary research can result in more holistic and efficient farming practices. Such as evaluating the environmental impact of smart irrigation, including its effects on soil health, water quality, and biodiversity. This research can inform sustainable farming practices. In terms of Public Awareness and Education. The research can also be used to educate farmers and the general public about the benefits of smart water irrigation and how to effectively use these systems. Awareness campaigns can promote responsible water usage. Studying ways to make smart irrigation technology more affordable and accessible to small-scale farmers and growers.

This research can bridge the technology gap and promote inclusivity. Smart water irrigation has the potential to revolutionize agriculture, enhance environmental sustainability, and address pressing global challenges related to water resources and food security. These implications highlight the importance of continued research and innovation in this field.

## ACKNOWLEDGEMENT

The researchers would like to express our heartfelt gratitude to God for His unwavering guidance, strength, and inspiration throughout this research. Without His divine wisdom and blessings, this work would not have been possible. We are humbled and thankful for the spiritual support that has sustained us during this journey.

## FUNDING

The study did not receive funding from any institution.

# DECLARATIONS

## *Conflict of Interest*

The authors declared that there is no interest in conflict associated with this research.

## *Informed Consent*

No direct, private personal information was used in the conduct of this research

## *Ethics Approval*

As no private and personal information was used in the research, ethics approval is not necessary.

## Authors' Biography

Regie P. Binayao is a Bachelor of Science in Information Technology student at Northern Bukidnon State College, specializing in hardware and the Internet of Things.

Paul Vincent L. Mantua is a Bachelor of Science in Information Technology student at Northern Bukidnon State College, specializing in programming, hardware, and the Internet of Things.

Holy Rose May Namocatcat is a Bachelor of Science in Information Technology senior student at Northern Bukidnon State College, specializing in human-computer interaction, hardware, and the Internet of Things.

Jade Kachel Klient Seroy is a Bachelor of Science in Information Technology student at Northern Bukidnon State College, specializing in hardware and the Internet of Things.

Phoebe Ruth Alithea B. Sudaria is an instructor at Northern Bukidnon State College, specializing in Website Development, Management Information Systems, Programming, the Internet of Things, and hardware implementations.

Kenn Migan Vincent C. Gumonan is an instructor for the Bachelor of Science in Information Technology at Cebu Institute of Technology University specializing in research, programming, game development, the Internet of Things, and project management.

Shiela Mae M. Orozco is the program head for the Bachelor of Science in Information Technology at Northern Bukidnon State College, specializing in programming, game development, Graphics design, Web development, and project management.